\documentclass[a4paper,11pt,twocolumn,tightenlines,accepted=2020-03-02]{quantumarticle}

\pdfoutput=1

\usepackage[utf8]{inputenc}
\usepackage[english]{babel}
\usepackage[T1]{fontenc}
\usepackage[a4paper, margin=1in]{geometry}
\usepackage{graphicx}
\usepackage{amsmath}
\usepackage{amsfonts}
\usepackage{tikz}
\usetikzlibrary{decorations.markings}
\usetikzlibrary{matrix}
\usepackage{braids}
\usepackage{bm}
\usepackage{physics}
\usepackage[numbers,sort]{natbib}
\usepackage{hyperref}

\begin{document}

\title{A Hierarchy of Anyon Models Realised by Twists in Stacked Surface Codes}
\author{T. R. Scruby}\email{thomas.scruby.17@ucl.ac.uk}
\affiliation{Dept. of Physics and Astronomy, University College London, London WC1E 6BT, UK}
\author{D. E. Browne}
\affiliation{Dept. of Physics and Astronomy, University College London, London WC1E 6BT, UK}

\begin{abstract}
    Braiding defects in topological stabiliser codes can be used to fault-tolerantly implement logical operations. Twists are defects corresponding to the end-points of domain walls and are associated with symmetries of the anyon model of the code. We consider twists in multiple copies of the 2d surface code and identify necessary and sufficient conditions for considering these twists as anyons: namely that they must be self-inverse and that all charges which can be localised by the twist must be invariant under its associated symmetry. If both of these conditions are satisfied the twist and its set of localisable anyonic charges reproduce the behaviour of an anyonic model belonging to a hierarchy which generalises the Ising anyons. We show that the braiding of these twists results in either (tensor products of) the S gate or (tensor products of) the CZ gate. We also show that for any number of copies of the 2d surface code the application of H gates within a copy and CNOT gates between copies is sufficient to generate all possible twists.
\end{abstract}

\maketitle

\section{Introduction}
\label{intro}
The question of which quantum gates can be performed fault-tolerantly in a particular quantum error correcting code is of vital importance if we wish to use the code in quantum computation. The conventional method of achieving fault-tolerant operations is via the application of transversal gates \cite{nielsen_quantum_2011}. In topological codes these have been generalised to \textit{locality preserving logical operators} \cite{bravyi_classification_2013}\cite{webster_locality-preserving_2018} which map local errors to local errors. However, these codes also allow us to perform fault-tolerant gates using another method: braiding of topological defects. Examples of topological defects include punctures (produced by the removal of stabilisers) and twists (the end-points of domain walls in the code). Braiding of these defects can allow us access to gates which are difficult or impossible to implement transversally. For example, the S gate cannot be implemented transversally in the 2d surface code (outside of folded surface code constructions such as \cite{moussa_transversal_2016}) but can be implemented via braiding of twist defects \cite{brown_poking_2017}.

A recent paper by M. Kesselring et al \cite{kesselring_boundaries_2018} fully categorises the twists of the 2d colour code, sorting them into nine conjugacy classes. In light of this result it seems natural to ask what gates we can implement via the braiding of these twists. In this work we attempt to answer this question for at least some of the colour code twists. In \cite{brown_poking_2017} the fact that braiding twists produces an S gate is shown by considering the action of this braid on the logical operators of the code but the same result can be obtained by considering the twists as (Ising) anyons and analysing their braiding relations, as in \cite{bombin_topological_2010}. Formally the twists in a topological code are described by G-crossed braided tensor categories \cite{barkeshli_symmetry_2014} and cannot in general be considered as anyons. We will discuss below the cases in which neglecting the full G-crossed category treatment of these defects is permissable and we will see that three of the nine conjugacy classes of colour code twist are examples of cases where twists can be analysed using anyonic models. These models are members of a hierarchy of anyonic models generalising the standard Ising anyon model used to study twists in the surface code. 

This paper is organised as follows: in section \ref{background} we present a short overview of anyon fusion and braiding relations and twists in topological codes, touching briefly on the G-crossed braided tensor category formalism and the occasions when it is acceptable to disregard it. We then define a hierarchy of ``extended Ising models'' in section \ref{hierarchy} and discuss the general fusion and braiding relations for these models in sections \ref{F} and \ref{R} respectively. In section \ref{gates} we clarify the correspondence between these relations and the possible logical operations that can be performed using these anyons. Finally, in section \ref{stack} we discuss how general models in this hierarchy can be realised in stacks of 2d surface codes with special attention given to the case of the 2d colour code. 

\section{Anyons and Twists}
\label{background}
In this section we briefly review the theoretical background necessary for the rest of the paper. We assume that readers are familiar with topological stabiliser codes and so our focus is on providing an outline of anyon fusion and braiding relations in sections \ref{anyons} and \ref{anyon_examples}. We refer readers who are not familiar with these codes to references \cite{gottesman_stabilizer_1997,bombin_topological_2006,kubica_abcs_2018,bombin_introduction_2013,fowler_surface_2012}. In section \ref{twists} we present a similar discussion regarding twists in topological codes and briefly touch on the category theory formalism that describes these objects. 

\subsection{Fusion and Braiding}
\label{anyons}

There exist a wide variety of ways to describe anyon models. They can be described in terms of topological charges \cite{bombin_topological_2010}, unitary braided tensor categories \cite{barkeshli_symmetry_2014} and through the lens of conformal field theory \cite{bais_condensate_2009} \cite{francesco_conformal_1997}. For our purposes the topological charge description is largely sufficient, although we will very briefly use the category-theoretic approach when discussing twists in section \ref{twists}. This means that we have some finite number of anyonic species, each possessing a unique label and topological charge. The anyons obey a set of fusion and braiding relations. Fusion relations are generally written in the form \cite{pachos_introduction_2012}
\begin{equation}
    a \times b = \sum_cN_{ab}^cc
\end{equation}

\noindent where $N_{ab}^c$ is an integer counting the number of ways anyons $a$ and $b$ can fuse into $c$. For a given charge $a$ if for any charge $b$ $N_{ab}^c$ is non-zero for at most a single charge $c$, in other words the fusion of $a$ with any other anyon has only one possible result, then we say that $a$ is Abelian. Otherwise it is non-Abelian. All anyon models must contain a unique vacuum charge ${\bf1}$ such that $a \times {\bf1} = a$. Additionally, each charge $a$ in the model must have a unique inverse $\bar{a}$ with which it can fuse to the vacuum in a unique way ($N_{a\bar{a}}^{\bf1} = 1$).

The total anyonic charge within a given region is a topological invariant and so cannot be altered by operations within this region. However, through alterations to anyon fusion order and intermediate fusion outcomes we can arrive at this same total charge via different paths, called fusion channels. This gives rise to the notion of an anyonic fusion space with dimension equal to the number of possible fusion channels. The quantum dimension of an anyon is defined to be 
\begin{equation}
    d_a^2 = \sum_b N_{a\bar{a}}^bd_b
\end{equation}

\noindent meaning that $d_a = 1$ for all Abelian anyons and $d_a > 1$ for non-Abelian anyons. The dimension of the fusion space of $N$ anyons $a$ grows asymptoticlly as $(d_a)^N$ in the limit of large $N$. Clearly if we wish to use anyons in quantum computation then only those models which contain non-Abelian anyons are of interest to us.

Changes of basis in the fusion space can be described using F-moves 

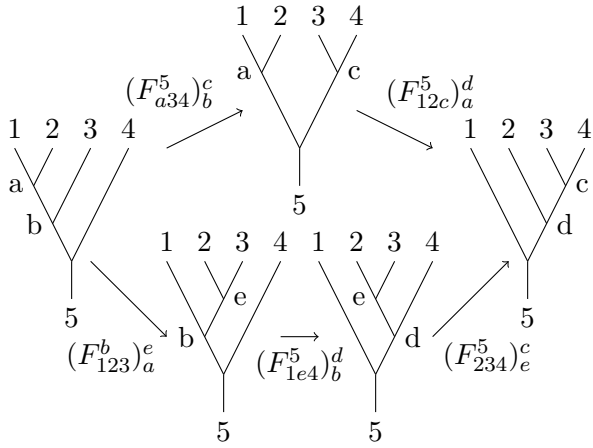
\begin{figure}
    \begin{center}
        \begin{tikzpicture}[scale=0.5]
            \draw (0,6) -- (1.5,3);
            \node [above] at (0,6) {1};
            \draw (1,6) -- (0.5,5);
            \node [above] at (1,6) {2};
            \node [left] at (0.5,5) {a};
            \draw (2,6) -- (1,4);
            \node [above] at (2,6) {3};
            \node [left] at (1,4) {b};
            \draw (3,6) -- (1.5,3);
            \node [above] at (3,6) {4};
            \draw (1.5,3) -- (1.5,2);
            \node [below] at (1.5,2) {5};

            \draw [->] (4,6) -- (6,7);
            \node [left] at (5.5,7.5) {$(F_{a34}^5)_b^c$};
            
            \draw [->] (2,3) -- (4,1);
            \node [left] at (4,0.5) {$(F_{123}^b)_a^e$};

            \draw (6,9) -- (7.5,6);
            \node [above] at (6,9) {1};
            \draw (7,9) -- (6.5,8);
            \node [above] at (7,9) {2};
            \node [left] at (6.5,8) {a};
            \draw (8,9) -- (8.5,8);
            \node [above] at (8,9) {3};
            \node [right] at (8.5,8) {c};
            \draw (9,9) -- (7.5,6);
            \node [above] at (9,9) {4};
            \draw (7.5,6) -- (7.5,5);
            \node [below] at (7.5,5) {5};

            \draw [->] (9,7) -- (11,6);
            \node [right] at (9.5,7.5) {$(F_{12c}^5)_a^d$};
                        
            \draw (4,3) -- (5.5,0);
            \node [above] at (4,3) {1};
            \draw (5,3) -- (5.5,2);
            \node [above] at (5,3) {2};
            \node [right] at (5.5,2) {e};
            \draw (6,3) -- (5,1);
            \node [above] at (6,3) {3};
            \node [left] at (5,1) {b};
            \draw (7,3) -- (5.5,0);
            \node [above] at (7,3) {4};
            \draw (5.5,0) -- (5.5,-1);
            \node [below] at (5.5,-1) {5};

            \draw [->] (7,1) -- (8,1);
            \node [below] at (7.5,1) {$(F_{1e4}^5)_b^d$};
            
            \draw (8,3) -- (9.5,0);
            \node [above] at (8,3) {1};
            \draw (9,3) -- (10,1);
            \node [above] at (9,3) {2};
            \node [right] at (10,1) {d};
            \draw (10,3) -- (9.5,2);
            \node [left] at (9.5,2) {e};
            \node [above] at (10,3) {3};
            \draw (11,3) -- (9.5,0);
            \node [above] at (11,3) {4};
            \draw (9.5,0) -- (9.5,-1);
            \node [below] at (9.5,-1) {5};

            \draw [->] (11,1) -- (13,3);
            \node [right] at (11,0.5) {$(F_{234}^5)_e^c$};
            
            \draw (12,6) -- (13.5,3);
            \node [above] at (12,6) {1};
            \draw (13,6) -- (14,4);
            \node [above] at (13,6) {2};
            \node [right] at (14,4) {d};
            \draw (14,6) -- (14.5,5);
            \node [right] at (14.5,5) {c};
            \node [above] at (14,6) {3};
            \draw (15,6) -- (13.5,3);
            \node [above] at (15,6) {4};
            \draw (13.5,3) -- (13.5,2);
            \node [below] at (13.5,2) {5};

        \end{tikzpicture}
    \caption{Diagrammatic representation of the pentagon equation. Different sequences of F-moves that have the same start and end point must be equivalent.}    
    \label{pentagon}
    \end{center}
\end{figure}

\begin{center}
    \begin{tikzpicture}

        \draw (0,3) -- (1,1);
        \node [above] at (0,3) {a};
        \draw (1,3) -- (0.5,2);
        \node [above] at (1,3) {b};
        \node [left] at (0.5,2) {u};
        \draw (2,3) -- (1,1);
        \node [above] at (2,3) {c};
        \draw (1,1) -- (1,0);
        \node [below] at (1,0) {d};

        \node at (2.2,1) {=};
        \node [right] at (2.5,1) {$\sum_v(F_{abc}^d)_u^v$};

        \draw (4,3) -- (5,1);
        \node [above] at (4,3) {a};
        \draw (5,3) -- (5.5,2);
        \node [above] at (5,3) {b};
        \node [right] at (5.5,2) {v};
        \draw (6,3) -- (5,1);
        \node [above] at (6,3) {c};
        \draw (5,1) -- (5,0);
        \node [below] at (5,0) {d};

    \end{tikzpicture}
\end{center}

\noindent where $u$ ($v$) can be any of the fusion outcomes of $a$ and $b$ ($b$ and $c$). More generally we should include additional indices describing the precise fusion channel by which $a$ and $b$ ($b$ and $c$) fuse to $u$ ($v$) but in what follows we will only consider fusion rules with $N_{ab}^c$ equal to either 1 or 0 and so such generality is unnecessary. The F-matrices associated with an anyon model can be found from the fusion rules by solving the pentagon equation \cite{pachos_introduction_2012}
\begin{equation}
    \label{pentagoneq}
    (F_{12c}^5)_a^d(F_{a34}^5)_b^c = \sum_e(F_{234}^d)_e^c(F_{1e4}^5)_b^d(F_{123}^b)_a^e
\end{equation}

\noindent which can be understood diagramatically in Fig. \ref{pentagon}.

Logical operations on the fusion space can be performed via the braiding of anyon pairs. This is an operation

\begin{center}
    \begin{tikzpicture}
    
        \node at (0,1) {$R_{ab}^c$};

        \draw (0,3) -- (0.5,1);
        \node [above] at (0,3) {a};
        \draw (1,3) -- (0.5,1);
        \node [above] at (1,3) {b};
        \draw (0.5,1) -- (0.5,0);
        \node [below] at (0.5,0) {c};

        \node at (1.5,1) {=};
        
        \braid [number of strands=2] (braid) at (2,3) s_1;
        \node [above] at (2,3) {a};
        \node [above] at (3,3) {b};
        \draw (2,1.5) to [out=270,in=120] (2.5,1);
        \draw (3,1.5) to [out=270,in=60] (2.5,1);
        \draw (2.5,1) -- (2.5,0);
        \node [below] at (2.5,0) {c};

    \end{tikzpicture}
\end{center}

\noindent which exchanges the positions of anyons $a$ and $b$ which fuse to $c$. $R_{ab}^c$ will be a phase if $a$ and $b$ have only a single possible fusion outcome or a diagonal matrix indexed by $c$ if there are multiple possible outcomes. The charges $a$ and $b$ and the outcome of their fusion $c$ are left unchanged by the braiding in accordance with the fact that anyonic charges cannot be modified through local operations. However, the fusion outcome of $a$ or $b$ with a third anyon may be modified by this braid. If the F-matrices for a model are known then we can find the R-matrices for that model using the hexagon equation
\begin{equation}
    \label{hexagoneq}
    R_{13}^c(F_{213}^4)_a^cR_{12}^a = \sum_b(F_{231}^4)_b^cR_{1b}^4(F_{123}^4)_a^b.
\end{equation}

\begin{figure}
    \begin{center}
        \begin{tikzpicture}[scale=0.5]

            \braid [number of strands=3] at (0,8) s_2^{-1} s_1^{-1};
            \node [above] at (0,8) {2};
            \node [above] at (1,8) {3};
            \node [above] at (2,8) {1};
            \draw (0,5.5) to [out=270,in=120] (0.5,4.5) to [out=270,in=120] (1,3.5);
            \draw (1,5.5) to [out=270,in=60] (0.5,4.5);
            \node [left] at (0.5,4.5) {a};
            \draw (2,5.5) to [out=270,in=60] (1,3.5);
            \draw (1,3.5) -- (1,2.5);
            \node [below] at (1,2.5) {4};

            \draw [->] (2.5,8) -- (3.5,9);
            \node [right] at (2.5,7.5) {$R_{12}^a$};

            \draw [->] (2.5,4) -- (3.5,3);
            \node [left] at (4,2) {$(F_{123}^4)_a^b$};

            \braid [number of strands=3] at (4,12) s_2^{-1};
            \node [above] at (4,12) {2};
            \node [above] at (5,12) {3};
            \node [above] at (6,12) {1};
            \draw (4,10.5) -- (4,9.5);
            \draw (5,10.5) -- (5,9.5);
            \draw (6,10.5) -- (6,9.5);
            \draw (4,9.5) to [out=270,in=120] (4.5,8.5) to [out=270,in=120] (5,7.5);
            \draw (5,9.5) to [out=270,in=60] (4.5,8.5);
            \node [left] at (4.5,8.5) {a};
            \draw (6,9.5) to [out=270,in=60] (5,7.5);
            \draw (5,7.5) -- (5,6.5);
            \node [below] at (5,6.5) {4};

            \draw [->] (6.5,9) -- (7.5,9);
            \node [below] at (7,9) {$(F_{213}^4)_a^c$};

            \braid [number of strands=3] at (8,12) s_2^{-1} 1;
            \node [above] at (8,12) {2};
            \node [above] at (9,12) {3};
            \node [above] at (10,12) {1};
            \draw (8,10.5) -- (8,9.5);
            \draw (9,10.5) -- (9,9.5);
            \draw (10,10.5) -- (10,9.5);
            \draw (8,9.5) to [out=270,in=120] (9,7.5);
            \draw (9,9.5) to [out=270,in=120] (9.5,8.5);
            \node [right] at (9.5,8.5) {c};
            \draw (10,9.5) to [out=270,in=60] (9.5,8.5) to [out=270,in=60] (9,7.5);
            \draw (9,7.5) -- (9,6.5);
            \node [below] at (9,6.5) {4};

            \draw [->] (10.5,9) -- (11.5,8);
            \node [left] at (11.5,7.5) {$R_{13}^c$};
                
            \braid [number of strands=3] at (4,4) s_2^{-1} s_1^{-1};
            \node [above] at (4,4) {2};
            \node [above] at (5,4) {3};
            \node [above] at (6,4) {1};
            \draw (4,1.5) to [out=270,in=120] (5,-0.5);
            \draw (5,1.5) to [out=270,in=120] (5.5,0.5);
            \draw (6,1.5) to [out=270,in=60] (5.5,0.5) to [out=270,in=60] (5,-0.5);
            \node [right] at (5.5,0.5) {b};
            \draw (5,-0.5) -- (5,-1.5);
            \node [below] at (5,-1.5) {4};

            \draw [->] (6.5,3) -- (7.5,3);
            \node [below] at (7,2) {$R_{1b}^4$};

            \draw (8,4) -- (8,1.5) to [out=270,in=120] (8.5,0.5) to [out=270,in=120] (9,-0.5);
            \draw (9,4) -- (9,1.5) to [out=270,in=60] (8.5,0.5);
            \draw (10,4) -- (10,1.5) to [out=270,in=60] (9,-0.5);
            \node [above] at (8,4) {2};
            \node [above] at (9,4) {3};
            \node [above] at (10,4) {1};
            \node [left] at (8.5,0.5) {b};
            \draw (9,-0.5) -- (9,-1.5);
            \node [below] at (9,-1.5) {4};

            \draw [->] (10.5,3) -- (11.5,4);
            \node [right] at (10,2) {$(F_{231}^4)_b^c$};

            \draw (12,8) -- (12,5.5) to [out=270,in=120] (13,3.5);
            \draw (13,8) -- (13,5.5) to [out=270,in=120] (13.5,4.5);
            \draw (14,8) -- (14,5.5) to [out=270,in=60] (13.5,4.5) to [out=270,in=60] (13,3.5);
            \node [above] at (12,8) {2};
            \node [above] at (13,8) {3};
            \node [above] at (14,8) {1};
            \node [right] at (13.5,4.5) {c};
            \draw (13,3.5) -- (13,2.5);
            \node [below] at (13,2.5) {4};
                
        \end{tikzpicture}
        \caption{Diagrammatic representation of the hexagon equation. As with the pentagon equation, different sequences of F and R-moves with the same start and end points must be equivalent}
        \label{hexagon}
    \end{center}
\end{figure}
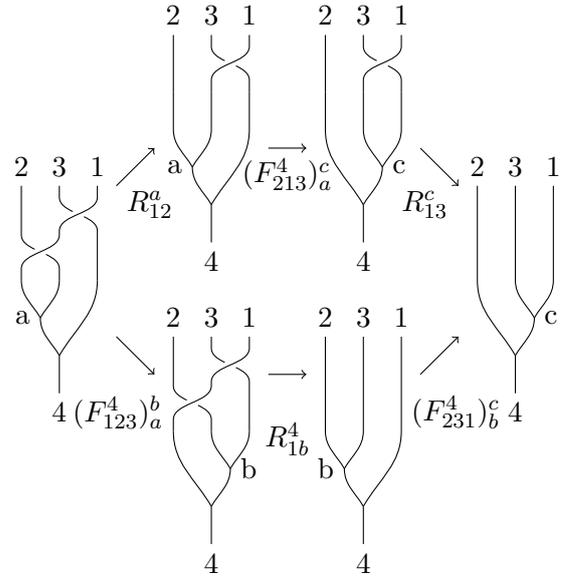

As with the pentagon equation, the hexagon equation can more easily be understood when presented diagramatically as in Fig. \ref{hexagon}.

\subsection{Examples}
\label{anyon_examples}

Two anyon models of central importance in our work are the quantum double of $\mathbb{Z}_2$ \cite{bombin_topological_2010} and the Ising anyons \cite{pachos_introduction_2012}. The former is an Abelian model with charges ${\bf1},e,m$ and $\epsilon$, fusion rules 
\begin{equation}
    \begin{split}
        & e \times e = m \times m = \epsilon \times \epsilon = {\bf1}, \\
        e \times &m = \epsilon, ~~ e \times \epsilon = m, ~~ m \times \epsilon = e
    \end{split}
\end{equation}

\noindent and braiding relations
\begin{equation}
    \begin{split}
        & R_{ee} = R_{mm} = 1, ~~ R_{\epsilon\epsilon} = -1, \\
        R_{em}&R_{me} = R_{e\epsilon}R_{\epsilon e} = R_{m\epsilon}R_{\epsilon m} = -1.
    \end{split}
\end{equation}

This model describes the excitations that arise in the toric code, with $e$ and $m$ anyons corresponding to $X$ and $Z$ errors and $\epsilon$ corresponding to a combination of the two (i.e. a $Y$ error). It also possesses a symmetry: we can exchange the $e$ and $m$ charge labels without affecting any of the fusion or braiding relations. 

In contrast, the Ising anyon model is non-Abelian. It contains three charges: ${\bf1},\psi$ and $\sigma$. The Ising anyon fusion rules are
\begin{equation}
    \psi \times \psi = {\bf1}, ~~ \psi \times \sigma = \sigma, ~~ \sigma \times \sigma = {\bf1} + \psi
\end{equation}

\noindent and the F matrix for the fusion of three $\sigma$s is
\begin{equation}
    F_{\sigma\sigma\sigma}^{\sigma} = \frac{1}{\sqrt{2}}
    \begin{pmatrix}
        1 & 1 \\
        1 & -1
    \end{pmatrix}
\end{equation}
                                                
\noindent and all other $F_{abc}^d$ are arbitrary phases. The braiding relations are
\begin{equation}
    \begin{split}
        & R_{\sigma\sigma} = e^{-i\pi/8}
        \begin{pmatrix}
            1 & 0 \\
            0 & i 
        \end{pmatrix}
        \\
        R_{\psi\psi} &= -1, ~~ R_{\psi\sigma}R_{\sigma\psi} = -1.
    \end{split}
\end{equation}

\subsection{Twists in Topological Codes}\label{twists}

\begin{figure}
    \begin{center}
        \includegraphics[width=.48\textwidth]{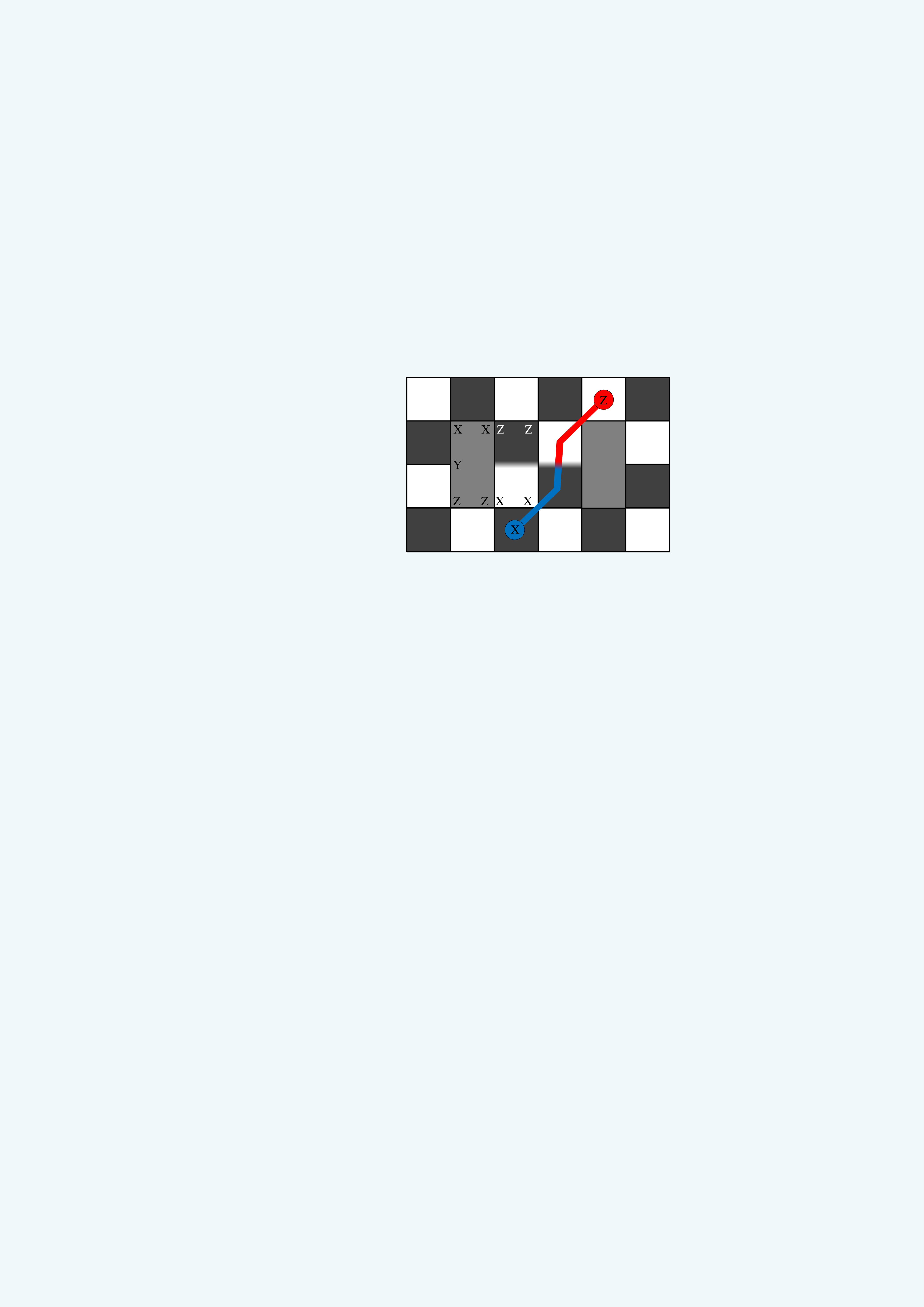}
        \caption{A domain wall and pair of twists in the 2d surface code. Qubits are on the vertices. Black plaquettes correspond to $Z$ stabilisers and white plaquettes to $X$ stabilisers. Each twist corresponds to a weight five stabiliser which includes one $Y$. $X$ and $Z$ stabilisers and excitations are exchanged by the domain wall.}
        \label{twistpic}
    \end{center}
\end{figure}

We noted above that the quantum double of $\mathbb{Z}_2$ is symmetric under exchange of $e$ and $m$ charges. We can consider a domain wall which applies precisely this symmetry, achievable in the toric code via a line of modified stabilisers each containing two $Z$ and two $X$ operators as shown in Fig. \ref{twistpic} \cite{brown_poking_2017} \cite{bombin_topological_2010}. An $X$ error moved across such a domain wall will be transformed to a $Z$ error and vice versa. The end points of domain walls such as this are called twists and are formally described by G-crossed braided tensor categories \cite{barkeshli_symmetry_2014}. We now give a very brief outline of some of the basic ideas of this formalism. This will be limited to the minimum details required for drawing a connection between twists and anyons and readers interested in a rigourous mathematical description of this formalism should refer to the sources cited.

\begin{figure}
    \begin{center}
        \begin{tikzpicture}
              
            \node [scale=1.2] at (-1.5,1) {a)};

            \draw [decoration={markings, mark = at position 0.25 with {\arrow[scale=2]{>}}},postaction={decorate}] (0,0) circle (1);
            \draw (0.25,0.25) rectangle (-0.25,-0.25);
            \node at (0,0) {$g$}; 
            \draw [dashed] (0,-.25) -- (0,-1.25);
            \draw [fill=black] (-1,0) circle (0.1);
            \node [left] at (-1,0) {$a_0$};

            \node [scale=1.2] at (2.5,1) {b)};
            
            \draw [decoration={markings, mark = at position 0.25 with {\arrow[scale=2]{>}}},postaction={decorate}] (4,0) circle (1);
            \draw (4.25,0.25) rectangle (3.75,-0.25);
            \node at (4,0) {$g$}; 
            \draw [dashed] (4,-.25) -- (4,-1.25);
            \draw [fill=black] (3,0) circle (0.1);
            \node [left] at (3,0) {$a_0$};
            \draw [fill=black] (4,0.5) circle (0.1);
            \node [left] at (4,0.5) {$b_0$};

        \end{tikzpicture}
        \caption{Braiding an anyon in a loop around a twist to measure the enclosed topological charge. $a_0$ is invariant under the symmetry $g$. cases a) and b) are distinguishable only if $a_0$ braids non-trivially with $b_0$.}
        \label{wilson}
    \end{center}
\end{figure}
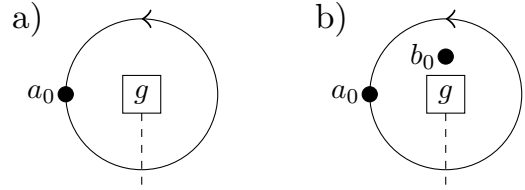

An anyon model can be described by a unitary braided tensor category $\mathcal{C}_0$ which has charges $a_0$ and a (possibly trivial) symmetry group $G$. The elements of $G$ are labelled $g$ and correspond to the symmetries of the anyon model. The identity element of this group is labelled $0$. The action of $g$ on $\mathcal{C}_0$ is an invertible map from $\mathcal{C}_0$ to itself. In a physical realisation of this anyon model each $g$ will correspond to a twist and braiding an anyon around this twist will apply the symmetry $g$ to that anyon,  with this action denoted as $^ga_0$. The topological charge of a twist can be measured by braiding it with an anyon $a_0$ which is invariant under the symmetry $g$ as in Fig. \ref{wilson} a). For each symmetry $g$ we have a new category $\mathcal{C}_{g}$ which has charges $a_{g}$. The number of distinct $a_g$ is equal to the number of $g$-invariant charges in $\mathcal{C}_0$. When fusing charges $a_{g}$ and $b_{h}$ we must have that $a_{g} \times b_{h} = c_{g\cdot h}$ where $g \cdot h$ is the composition of elements of $G$. This is called $G$-graded fusion.

Since $g \cdot 0 = g$ we have that $a_g \times b_0 = a'_g$. If the $g$-invariant charge(s) in $\mathcal{C}_0$ cannot distinguish $b_0$ from the vacuum then $a'_g = a_g$ and we say that anyon $b_0$ is ``localised'' by twist $a_g$. Such charges and twists have fusion/splitting rules such that 
\begin{equation}
    N_{a_{g}b_0}^{a_{g}} = N_{a_{g}\bar{b}_0}^{a_g} = N_{a_g\bar{a}_g}^{b_0} = N_{\bar{a}_gb_0}^{\bar{a}_g} = N_{^gb_0a_g}^{a_g} \neq 0
\end{equation}

In other words if $a_g$ localises $b_0$:

\begin{itemize}
    \item $\bar{a}_g$ also localises $b_0$.
    \item $b_0$ is one of the possible fusion outcomes of $a_g \times \bar{a}_g$.
    \item All charges in the orbit of $b_0$ under the action of $g$ are also localised by $a_g$ and $\bar{a}_g$.
\end{itemize}

Additionally we note that the set of localisable charges for a particular twist must be closed under fusion since if $a_0$ and $b_0$ both braid trivially with the $g$-invariant charges in $\mathcal{C}_0$ then so must the result of their fusion. 

Braiding of charges $a_g$ and $b_h$ involves the action of the relevant symmetries such that charges can be modified by these braids (this is the G-crossed braiding part of the formalism). However, we will not require this part of the theory for reasons that will become clear below.

Finally, we note that fusion rules in this formalism must still satisfy the pentagon equation. Braiding rules are generalised to follow a ``heptagon equation'' which accounts for the fact that braiding with twists can alter charge labels. 

We return now to the previously discussed case of twists in the toric code. $\mathcal{C}_0$ in this case is the quantum double of $\mathbb{Z}_2$ which has only one symmetry so $G$ has only two elements, $0$ and $g$, where $0$ is the identity. $\mathcal{C}_0$ contains two $g$-invariant charges (${\bf1}$ and $\epsilon$) so $\mathcal{C}_g$ has two charges which can be distinguished by braiding with $\epsilon$. More specifically there is one charge corresponding to the fusion of the twist with either ${\bf1}$ or $\epsilon$ and one charge corresponding to fusion with $e$ or $m$. This twist possesses two significant features: (1) it is self-inverse and (2) its associated invariant charges are also its localisable charges. This means that the subset of charges consisting of this twist and its localisable charges is closed under fusion. Furthermore, none of the charges in this subset can be altered by braiding with any of the others. In other words this subset functions as an anyon model - specifically the Ising anyon model. This is precisely what was noticed by H. Bombin in \cite{bombin_topological_2010}, although the argument in that paper was formulated in terms of topological string operators. 

In the 2d colour code the situation is not quite so simple. In \cite{kesselring_boundaries_2018} the 72 twists of the colour code are identified and arranged into nine conjugacy classes. The authors of \cite{kesselring_boundaries_2018} point out that the action of these twists can best be understood by considering the nine (non-trivial) bosonic anyons of the colour code arranged as in Table. \ref{ccbosons}. These anyons are all self-inverse and Abelian. Two anyons in a row or column fuse to the third and braid trivially with each other. Two anyons which do not share a row or column fuse to a fermion and aquire a phase of -1 under full exchange (monodromy).

\begin{table}
    \begin{center}
    \resizebox{.3\columnwidth}{!}{%
    \begin{tabular}{c|c|c}
        $rx$ & $gx$ & $bx$ \\
        \hline
        $ry$ & $gy$ & $by$ \\
        \hline
        $rz$ & $gz$ & $bz$
    \end{tabular}%
    }
    \caption{The nine non-trivial bosonic anyons of the 2d colour code arranged as in \cite{kesselring_boundaries_2018}. $rx$ implies an $X$ error on a red plaquette and so on.} 
    \label{ccbosons}
    \end{center}
\end{table}

The symmetries of the colour code anyon model are the permutations of this table which preserve the rows and columns. These permutations are the column permutations (6 options), row permutations (6 options) and the transpose (2 options) giving $6 \times 6 \times 2 = 72$ possible symmetries. Twists belonging to three of the nine conjugacy classes possess the same properties as the surface code twist described above: they are self-inverse and their associated sets of invariant and localisable charges are equivalent. One of the other six classes is trivial (it contains only the identity twist) and twists in the other five classes possess neither of these properties. 

In what follows we consider the general case of the anyon model associated with these self-inverse twists and their localisable charges. In section \ref{stack} we will show that twists in any number of stacked surface codes can only have an invariant set of localisable charges if they are self-inverse.

\section{A Hierarchy of Models}
\label{hierarchy}
Recall the standard Ising model in which we have a single non-Abelian anyon $\sigma$ and two Abelian anyons ${\bf1}$ and $\psi$ such that $\sigma \times \sigma = {\bf1} + \psi$. We can extend this model by including additional Abelian anyons and modifying the outcome of $\sigma \times \sigma$ to include these anyons. Such extended models already exist in the literature in the form of parafermions, for example in \cite{hutter_parafermions_2015}, but the Abelian anyons in these models are not generally self-inverse and so cannot be the natural anyons of the colour code. If we write the Abelian anyons of an extended Ising model as $\alpha_i$ (where $\alpha_0={\bf1}$) and the non-Abelian anyon as $\beta$ and require that each $\alpha_i$ must be its own antiparticle then we obtain the following fusion relations
\begin{equation}
    \label{frules}
    \alpha_i \times \alpha_j = \alpha_k, 
    ~~
    \alpha_i \times \beta = \beta, 
    ~~
    \beta \times \beta = \sum_{i=0}^n \alpha_i
\end{equation}

where $k=0$ only if $i=j$ and $k=i$ ($k=j$) only if $j=0$ ($i=0$). These are exactly the fusion relations we observe for self-inverse twists and their localisable anyons in the colour code. For example, the colour code twist $B$ is self-inverse and can localise the anyons $bx,by,bz$ from table \ref{ccbosons} (as well as the vacuum anyon). The set $\{\bm{1},bx,by,bz\}$ is closed under fusion and so if we write $\bm{1}=\alpha_0$, $bx=\alpha_1$, $by=\alpha_2$, $bz=\alpha_3$ and $B=\beta$ then fusion relations of these five charges exactly match (\ref{frules}).

Only specific values of $n$ yield valid extended Ising models. For example, the model $\{\alpha_0,\alpha_1,\alpha_2,\beta\}$ is not valid because it is not closed under fusion ($\alpha_1 \times \alpha_2$ must have a fusion outcome $\alpha_k$ where $k \neq 0,1,2$). Given a valid extended Ising model containing $m$ $\alpha$s we can find the next valid model with $n > m$ by adding a single new charge $\alpha_{m+1}$ to the model, fusing $\alpha_{m+1}$ with all existing charges, and adding all fusion outcomes to the model. If we write these fusion outcomes as $\alpha_i \times \alpha_{m+1} = \alpha_{i+m+1}$ then we can see that the resulting model must be closed under fusion since
\begin{itemize}
    \item Fusion of any $\alpha_i,\alpha_j$ with $i,j \leq m$ results in another $\alpha_k$ with $k \leq m$ since the initial model was closed under fusion.
    \item Fusion of $\alpha_{m+1}$ with any anyon in the model results in another anyon in the model due to the above procedure.
    \item Fusion of any $\alpha_i,\alpha_j$ with $i \leq m$ and $j \geq m+2$ can be written as $\alpha_{i \leq m} \times \alpha_{j \geq m+2} = \alpha_{i \leq m} \times \alpha_{k \leq m} \times \alpha_{m+1}$ by definition of $\alpha_{j \geq m+2}$, and this is in the model due to the above two points. 
    \item Fusion of any $\alpha_i, \alpha_j$ with $i,j \geq m+2$ can be written as $\alpha_{i \geq m+2} \times \alpha_{j \geq m+2} = \alpha_{k \leq m} \times \alpha_{m+1} \times \alpha_{l \leq m} \times \alpha_{m+1}$ which is in the model since the two $\alpha_{m+1}$s cancel. 
\end{itemize}

Thus we can inductively define all extended Ising models beginning from the standard Ising model: $\{\alpha_0,\alpha_1,\beta\}$. We can label these models by $I_k$ where $k=1$ is the standard Ising model. The number of $\alpha_i$ in a given $I_k$ is $n_k = 2n_{k-1} = 2^{k-1}n_1$ and $n_1 = 2$ so $n_k = 2^k$. The $\beta$ anyon for each $I_k$ can be written as $\beta_k$, and it has quantum dimension $\sqrt{2^k}$. 

Note that we can equivalently define these models from multiple copies of $\mathbb{Z}_2$. The Abelian charges of the standard Ising anyon model form a group (with composition of group elements described by the model's fusion rules) that is isomorphic to $\mathbb{Z}_2$. Similarly the Abelian charges of $I_2$ form a group that is isomorphic to $\mathbb{Z}_2 \times \mathbb{Z}_2$ and so on. In general the group of Abelian charges of $I_k$ will be isomorphic to $k$ copies of $\mathbb{Z}_2$. A set consisting of a finite group $A$ and a single additional element $\beta$ with composition of these elements defined as in (\ref{frules}) is called a Tambara-Yamagami (TY) category \cite{tambara_tensor_1998}. Thus our ``extended Ising hierarchy'' can be described more formally as a hierarchy of TY categories with categories in the $k^{\textrm{th}}$ level of the hierarchy having base group $(\mathbb{Z}_2)^k$.

\section{F Matrices}
\label{F}
In this section we show the possible F matrices $F_{\beta\beta\beta}^{\beta}$ for general $\beta_k$. The full derviation of these matrices can be found in appendix \ref{Fapp}. These matrices (up to symmetry-preserving row and column permutations) are found to be a subset of the Hadamard matrices called Sylvester or Walsh matrices \cite{sylvester_lx._1867} multiplied by a constant. The implications of this fact on the trace of these matrices is examined as this will be relevant in section \ref{R}. 

A general F-matrix for extended Ising anyons can be found from the pentagon equation (\ref{pentagoneq}). Every $F_{abc}^d$ is equal to zero (if it is disallowed by the fusion rules) or a phase except for $F_{\beta\beta\beta}^{\beta}$ which involves only non-Abelian anyons. In appendix \ref{Fapp} we use the pentagon equation to show that up to a choice of gauge this matrix has the form
\begin{equation}
    F_{\beta\beta\beta}^{\beta} = \pm\frac{1}{\sqrt{2^k}}\bm{\phi}
\end{equation}

\noindent where $\bm{\phi}$ is a symmetric Hadamard matrix with elements given by 

\begin{equation}
    \label{phi}
    \phi_{ij} = (F_{\beta\alpha_i\beta}^{\alpha_j})_{\beta}^{\beta} = (F_{\alpha_i\beta\alpha_j}^{\beta})_{\beta}^{\beta}.
\end{equation}

Hadamard matrices are $n\times n$ matrices where all entries are $\pm1$ and $\bm{MM}^T=n\bm{I}$ \cite{hedayat_hadamard_1978}. Additionally, $\bm{\phi}$ as we have defined it has the property that all entries in the first row and column are $+1$ (such Hadamard matrices are called ``normalised'' but we will avoid using this term to prevent confusion with its more common usage in quantum physics). We can further restrict $\bm{\phi}$ to a subset of Hadamard matrices by using the group theory and TY category connections discussed previously. The F matrices of TY categories are related to symmetric non-degenerate bicharacters where ``character'' refers to multiplicative character, i.e. a group homomorphism from a group $A$ to the multiplicative group of a field $\bm{F}^\times$ \cite{emil_artin_galois_1959}. A bicharacter is then a function $\chi: A \times A \rightarrow \bm{F}^\times$ which satisfies \cite{rozanski_bicharacters_1996}
\begin{equation}
    \begin{split}
        &\chi(a_1a_2,a_3) = \chi(a_1,a_3)\chi(a_2,a_3) \\
        &\textrm{and} \\
        &\chi(a_1,a_2a_3) = \chi(a_1,a_2)\chi(a_1,a_3).
    \end{split}
\end{equation}

We can see that $\bm{\phi}$ satisfies this by considering the pentagon equation with $1=3=\beta$, $2=\alpha_i$, $4=\alpha_j$ and $5=\alpha_k$ which yields a constraint
\begin{equation}
    (F_{\beta\alpha_i\beta}^{\alpha_k})_{\beta}^{\beta} = (F_{\alpha_i\beta\alpha_j}^{\beta})_{\beta}^{\beta}
                                                          (F_{\beta\alpha_i\beta}^{(\alpha_j \times \alpha_k)})_{\beta}^{\beta}
\end{equation}

\noindent which can be rewritten as 
\begin{equation}
    \phi_{i(j \times k)} = \phi_{ij}\phi_{ik}
\end{equation}

\noindent using (\ref{phi}) and the fact that $(\phi_{ij})^{-1} = \phi_{ij}$ since $\phi_{ij} = \pm1$. $\bm{\phi}$ is symmetric and so we also have that
\begin{equation}
    \phi_{(j \times k)i} = \phi_{ji}\phi_{ki}.
\end{equation}

Thus $\bm{\phi}$ is a bicharacter on $(\mathbb{Z}_2)^k$. The above definition means that if we fix one argument of a bicharacter on a group $A$ then the bicharacter as a function of the other argument defines a character on $A$. In other words, each row and column of $\bm{\phi}$ is a character on $(\mathbb{Z}_2)^k$. Because they are homomorphisms the action of these characters on $(\mathbb{Z}_2)^k$ is defined by their action on each of the $k$ copies of $\mathbb{Z}_2$ which make it up. There are two valid $\pm1$ valued characters on $\mathbb{Z}_2$ which are given by the rows/columns of the $2\times2$ Hadamard matrix
\begin{equation}
    \label{H1}
    H_1 = 
    \begin{pmatrix}
        1 & 1 \\
        1 & -1
    \end{pmatrix}
\end{equation}

\noindent and coincide with the irreducible representations of $\mathbb{Z}_2$. Thus for $(\mathbb{Z}_2)^k$ there are $2^k$ possible characters corresponding to the rows/columns of $H_1^{\otimes k}$. These matrices are a subset of the Hadamard matrices called Sylvester or Walsh matrices \cite{sylvester_lx._1867}. The possible bicharacters for $(\mathbb{Z}_2)^k$ are then the Sylvester matrix $H_1^{\otimes k}$ and any other matrices which can be obtained from this matrix via symmetry-preserving row/column permutations.

$H_1$ is the unique bicharacter for $k=1$. For $k=2$ there are four possible bicharacters which we can write in matrix form as 
\begin{equation}
    \begin{split}
        &\begin{pmatrix}
            1 & 1 & 1 & 1 \\
            1 & 1 & -1 & -1 \\
            1 & -1 & 1 & -1 \\
            1 & -1 & -1 & 1 
        \end{pmatrix}
        ~~
        \begin{pmatrix}
            1 & 1 & 1 & 1 \\
            1 & -1 & 1 & -1 \\
            1 & 1 & -1 & -1 \\
            1 & -1 & -1 & 1
        \end{pmatrix}
        \\
        &\begin{pmatrix}
            1 & 1 & 1 & 1 \\
            1 & -1 & -1 & 1 \\
            1 & -1 & 1 & -1 \\
            1 & 1 & -1 & -1 
        \end{pmatrix}
        ~~
        \begin{pmatrix}
            1 & 1 & 1 & 1 \\
            1 & 1 & -1 & -1 \\
            1 & -1 & -1 & 1 \\
            1 & -1 & 1 & -1 
        \end{pmatrix}.
    \end{split}
    \label{k2fmat}
\end{equation}

One of these matrices has trace 4 while the other three have trace 0. These three all correspond to the same anyon model up to relabelling of charges. In appendix \ref{permuteapp} we show that symmetry-preserving permutations of columns of a symmetric Hadamard matrix must alter the trace of the matrix by either 0 or $\pm2^k$, with the latter only being possible for even $k$. The Sylvester matrices all have trace 0 and since one row/column contains only $+1$s we cannot have Tr$(\bm{\phi})=-2^k$ so the only possible traces are 0 for odd $k$ and $0,2^k$ for even $k$. In comparison general symmetric $2^k\times2^k$ Hadamard matrices must also have trace 0 for odd $k$ but the trace can take any value $2^k - 2^{k/2+1} \geq 0$ for even $k$ \cite{craigen_trace_1994}. 

\section{R Matrices}
\label{R}
We now discuss the possible R matrices for extended Ising models which are found from the hexagon equation (\ref{hexagoneq}). This equation was solved for the case of TY categories in \cite{siehler_braided_2000}. Once again a similar solution in the language of section \ref{background} can be found in appendix \ref{Rapp}. 

We are concerned only with $R_{\beta\beta}$ which are diagonal matrices. From the hexagon equation we obtain the constraint
\begin{equation}
    \label{Rsec1}
    R_{\beta\beta}^{\alpha_i} = \pm\sqrt{\phi_{ii}}R_{\beta\beta}^{\alpha_0}
\end{equation}

\noindent with $R_{\beta\beta}^{\alpha_0}$ having possible values $\{\pm1,\pm i,\pm e^{i\pi/4},\pm e^{-i\pi/4}\}$ for even $k$ and $\{\pm e^{i\pi/8},\pm e^{-i\pi/8},\pm ie^{i\pi/8},\pm ie^{-i\pi/8}\}$ for odd $k$. This equation tells us that (up to a global phase) the non-zero elements of $R$ are equal to $\pm 1$ and $\pm i$, with the number of real (imaginary) elements equal to the number of $+1$s ($-1$s) in the diagonal of $\bm{\phi}$. 

The hexagon equation also gives a constraint on the trace of $R$, 

\begin{equation}
    \label{Rsec2}
    \pm \frac{R_{\beta\beta}^{\alpha_0}\textrm{Tr}(R)}{\sqrt{2^k}} = 1,
\end{equation}

\noindent which tells us about the relative numbers of positive and negative diagonal elements. We can use (\ref{Rsec1}) to rewrite this as

\begin{equation}
    \label{Rsec3}
    \pm \frac{(R_{\beta\beta}^{\alpha_0})^2(a+ib)}{\sqrt{2^k}} = 1
\end{equation}

\noindent where $a$ and $b$ are integers. In order for $|R_{\beta\beta}^{\alpha_0}|^2=1$ we need $a^2 + b^2 = 2^k$. For even $k$ the solutions to this equation are $a = \pm 2^{k/2},b=0$ and $a=0,b = \pm 2^{k/2}$ and for odd $k$ they are $a = \pm b = \pm 2^{(k-1)/2}$ (see appendix \ref{Rapp}). Combined with (\ref{Rsec1}) this completely determines the number of $+1$, $-1$, $+i$ and $-i$ elements on the diagonal of $R_{\beta\beta}$ for a given $k$ and $\textrm{Tr}(\bm{\phi})$ up to global phase, with the number of each shown for each case in table \ref{Rtab}.

\begin{table}
    \begin{tabular}{c c c c}
        \hline
        $k$ & Tr$(\bm{\phi})$ & Number of $\pm 1$ & Number of $\pm i$ \\
        \hline
        Even & $2^k$ & $2^{k-1}\pm2^{k/2-1}$ & 0 \\
        Even & 0     & $2^{k-2}\pm2^{k/2-1}$ & $2^{k-2}$ \\
        Odd  & 0     & $2^{k-2}\pm2^{(k-3)/2}$ & $2^{k-2}\pm2^{(k-3)/2}$ \\
        \hline
    \end{tabular}
    \caption{The possible numbers of $\pm 1$ and $\pm i$ on the diagonal of $R_{\beta\beta}$ up to global phase. The $\pm$ signs in a given column correspond to the $\pm$ in that column's header such that choosing the sign in the header to be $+$ also fixes all $\pm$ in the column to $+$.}
    \label{Rtab}
\end{table}

We also show in appendix \ref{Rapp} that the elements $\phi_{ii}$ describe the exchange statistics of the anyons $\alpha_i$, with $\phi_{ii}=1$ indicating that $\alpha_i$ is bosonic and $\phi_{ii}=-1$ telling us that $\alpha_i$ is fermionic. Thus we expect that for both odd and even $k$ we can obtain models containing $2^{k-1}$ bosonic and $2^{k-1}$ fermionic Abelian charges and a single non-Abelian charge. We additionally expect that for even $k$ we can obtain models containing $2^k$ bosonic Abelian charges, no fermionic charges and a single non-Abelian charge. The models with $2^{k-1}$ bosonic and $2^{k-1}$ fermionic charges can be viewed as ``sub-models'' of $k$ copies of the standard Ising model (e.g. in some kind of multi-layer system) containing all Abelian charges and only a single non-Abelian charge (namely, the charge corresponding to $\sigma_1 \otimes \sigma_2 ... \otimes \sigma_k$). We note also that this ``sub-model'' simply corresponds to the case where we neglect some of the charges in the original model and does not mean that these charges are no longer present. It is therefore different from procedures such as that of Bais and Slingerland \cite{bais_condensate_2009} in which an actual change to the model is made. 

The models containing $2^k$ bosonic charges cannot be produced from copies of the standard Ising model and instead correspond to copies of a different anyonic model with four bosonic Abelian charges and a single non-Abelian charge. 

\section{Logical Gates}
\label{gates}
In this section we will rephrase the findings from the past two sections in terms of the possible logical gates which we can perform using these anyons.

Consider a specific anyon model containing $2^k$ Abelian charges and a single non-Abelian charge. This model has an $F$ matrix $F_{\beta\beta\beta}^{\beta}$ associated with changing the fusion order of three of the non-Abelian anyons and an $R$ matrix $R_{\beta\beta}$ associated with braiding two of the non-Abelian anyons. A system containing four such anyons has a $2^k$ dimensional fusion space for which the $2^k$ Abelian anyons of the model form a canonical basis. The $F$ and $R$ matrices provide us with two logical operations which can be performed on this space. The $F$ matrix is a mapping between the canonical basis and a basis of equal superpositions of the canonical basis vectors. The $R$ matrix selectively applies one of the phases $\{+1,-1,+i,-i\}$ to each vector of the canonical basis with the total number of $+1$s, $-1$s, $+i$s and $-i$s applied consistent with table \ref{Rtab}.

Both of these operations may be interpreted in terms of Clifford gates on $k$ qubits: the $F$ matrix has the same rows and columns as a tensor product of $k$ Hadamard gates (up to a global phase) and the same is true for the $R$ matrices and tensor products of either $k$ S gates or $k/2$ CZ gates. Notice that our canonical basis vectors are currently labelled only by anyonic charges and we have not yet defined an encoding for qubits in this space. We can always choose this encoding such that the ordering of the diagonal elements of $R$ in the computational basis is consistent with the respective tensor product of Clifford gates. The same is also true for the trace-0 $F$ matrices but not for the trace-$\sqrt{2^k}$ $F$ matrices since the trace is independent of our choice of encoding. These matrices cannot be decomposed into a tensor product of single qubit gates and instead correspond to tensor products of the trace-4 matrix in (\ref{k2fmat}) multiplied by $1/2$. This matrix is equivalent to SWAP$\cdot(H \otimes H)$ and so is also Clifford. Thus, up to global phases and a choice of encoding, all $F$ and $R$ matrices implement Clifford operations on our Hilbert space. 

\section{Stacked Surface Codes}
\label{stack}
Part of our motivation for obtaining the results presented in the previous sections was to examine the braiding relations of twists in the 2d colour code. In this section we will see that we can obtain twists belonging to the first and second levels of the hierarchy in this code and in general we can obtain twists belonging to the $k^{\textrm{th}}$ level of the hierarchy in a stack of $k$ 2d surface codes. When visualising such a stack we might imagine multiple copies of the surface code placed one above the other, but we allow operations between any two layers regardless of how ``far apart'' in the stack they are and thus there is no notion of distance in a third dimension and the entire stack is embeddable in 2d (although this will have an impact on the locality of the stabilisers).

The fact that models from the second level of the hierarchy can be realised in the 2d colour code is readily apparent from \cite{kesselring_boundaries_2018}. For example the twist $B$ which exchanges the $r$ and $g$ columns of Table \ref{ccbosons} has four nontrivial localisable charges: $\bm{1}$, $bx$, $by$ and $bz$. This set of four charges is closed under fusion and all four are invariant under the action of $B$. 

In general if we consider equivalence up to global phases and choose our qubit encoding as discussed in the previous section then there are two R-matrices for $k=2$:
\begin{equation}
    \label{CCMat2}
    \begin{pmatrix}
        1 & & & \\
        & 1 & & \\
        & & 1 & \\
        & & & -1 
    \end{pmatrix}
    ~~~~
    \begin{pmatrix}
        1 & & & \\
        & i & & \\
        & & i & \\
        & & & -1
    \end{pmatrix}
\end{equation}

Recall that the first of these matrices belongs to a model with four bosonic Abelian charges while the second belongs to a model with two bosonic and two fermionic charges. In the notation of Kesselring et al \cite{kesselring_boundaries_2018} the models containing four bosons are those associated with a twist from conjugacy class B while those containing two bosons and two fermions are associated with twists from conjugacy class C. The twists in conjugacy class G have only two localisable charges and correspond to models from the first level of the hierarchy.

So far we have seen that we can realise the $k=1$ level of the extended Ising hierarchy in the surface code, and the $k=2$ level in the colour code, which is equivalent to two copies of the surface code \cite{kubica_unfolding_2015}. We now consider the general case of a stack of $k$ copies of the surface code.

Consider the anyon model of $k$ stacked surface codes (in the absence of twists). The topological charges in this model are the elements of a finitely-generated free group, whose generating set can be written $\{e_1,m_1,e_2,m_2,...,e_k,m_k\}$ where the subscript shows the layer in the stack which the charge belongs to. Twists in the code stack correspond to symmetries of the anyon model. These symmetries can be formally defined as the elements of the automorphism group of the anyon model which preserve braiding relations. The action of these symmetries can be described via a set of orbits, each of which can be written as 
\begin{equation}
    a \rightarrow b \rightarrow c \rightarrow ... \rightarrow a
\end{equation}

\noindent with the ``trivial orbit'' defined as
\begin{equation}
    a \rightarrow a.
\end{equation}

We first show that only self-inverse symmetries $g$ of this anyon model can have a $g$-invariant set of localisable charges. Consider a twist $t_g$ and a charge $b_0 = a_0 \times {^ga_0}$. $b_0$ can be localised by $t_g$ because we can split it into $a_0$ and $^ga_0$, then braid $a_0$ around $t_g$ and fuse it to the vacuum with $^ga_0$. If we braid $b_0$ around $t_g$ we obtain ${^gb_0} = {^ga_0} \times {^{g^2}a_0}$ and so in order for $b_0$ to be $g$-invariant we must have ${^{g^2}a_0} = a_0$, implying that $g$ is self-inverse.  

From this we can see that if a twist in a stack of surface codes (together with its set of localisable charges) can be considered as an anyon model this model will belong to the hierarchy of extended Ising models defined in section \ref{hierarchy}. 

All non-trivial orbits associated with a self-inverse symmetry have the form $a \rightarrow b \rightarrow a$ which can also be written $a \leftrightarrow b$. 

The full automorphism group of a finitely generated free group with ordered basis $[x_1,...,x_n]$ can be generated by the \textit{elementary Neilsen transformations}\cite{magnus_combinatorial_2004}:

\begin{itemize}
    \item Switch $x_1$ and $x_2$
    \item Replace $x_1$ with $x_1^{-1}$
    \item Replace $x_1$ with $x_1 \cdot x_2$
\end{itemize}

The second transformation is equal to the identity transformation in our case because all charges in our model are their own inverse. We thus consider only the first and third transformations, but not all applications of these transformations are valid because we must also preserve braiding relations. In order to do this we require that if we map $x_i \rightarrow x_j$ then we must also map $x'_i \rightarrow x'_j$ and if we map $x_i \rightarrow x_ix_j$ then we must map $x'_j \rightarrow x'_ix'_j$, where $x_i$ can be either $e_i$ or $m_i$ and $x_ix'_i = \epsilon_i$. We also cannont map $e_i \rightarrow e_im_i$ within a layer because this exchanges a boson with a fermion. In other words all symmetries of the model can be generated by the transformations
\begin{equation} \label{gen1} e_i \leftrightarrow m_i \end{equation}
\begin{equation} \label{gen2} e_i \leftrightarrow e_j \textrm{ and } m_i \leftrightarrow m_j \end{equation}
\begin{equation} \label{gen3} e_i \leftrightarrow e_ie_j \textrm{ and } m_j \leftrightarrow m_im_j \end{equation}

    \noindent which are simply the generators of all colour code symmetries generalised to act on a stack of more than two surface codes \cite{kesselring_boundaries_2018}. A simple way to obtain a twist corresponding to a $\beta_k$ anyon is simply to combine twists associated with symmetry (\ref{gen1}) on $k$ different levels. The domain walls produced by these symmetries in the code correspond respectively to lines of H, SWAP and CNOT gates applied in the code stack. Since SWAP can be generated from CNOTs the generating set of symmetries can be reduced to just (\ref{gen1}) and (\ref{gen3}). This is consistent with the set of generating symmetries identified in \cite{webster_locality-preserving_2018} although we arrive at this result by a different method. Any product of these symmetries thus corresponds to a product of Clifford gates in the code stack and so the code containing the twists will also be a 2d stabiliser code. Braiding operations are performed using predefined sets of single-qubit Pauli measurements and additional modifications to stabilisers by the Clifford gates listed above \cite{brown_poking_2017} and such operations in a 2d stabiliser code should not result in a logical non-Clifford gate. Thus all twists produced by composition of these symmetries should have Clifford braiding relations.

This result is valid for more than just self-inverse twists since (\ref{gen1}-\ref{gen3}) are the generators of \textit{all} symmetries of the anyon model. Thus the restriction to Clifford braiding operations is valid for all twists in stacked surface codes. This is in agreement with recent results regarding the power of defect braiding in topological codes \cite{webster_braiding_2018}\cite{webster_fault-tolerant_2019}.

Finally we comment briefly on the fault-tolerance of such braiding procedures. As mentioned above, braiding operations with twists can be performed using the standard code deformation techniques of (1) measurement of modified stabilisers and (2) single-qubit Pauli measurements to remove physical qubits from the code and provide information for decoding \cite{brown_poking_2017}. In a code with local stabilisers these operations will also be local so we expect that braids with these generalised twists should remain fault-tolerant under existing decoding procedures. 

\section{Summary}
We have constructed a hierarchy of anyonic models extending the standard Ising anyon model and identified the significant properties of the F and R matrices for these models in the general case. These models are specific cases of Tambara-Yamagami categories and have $F$ and $R$ matrices belonging to the Clifford group. These anyon models can be realised using Abelian anyons and twists in stacked surface codes: given a stack of $k$ surface codes we can realise models belonging to level $k$ of the hierarchy. The restriction to Clifford group braiding relations of twists extends even to those twists which do not reproduce the behaviour of anyons. This result is consistent with other recent results in this area.

A number of possible future research directions exist following the results outlined above. Although we have shown that twists in stacked surface codes will always have Clifford braiding relations we have only characterised these relations for a small subset of these twists. Finding the braiding relations for the remaining twists will likely require the use of the full G-crossed braided tensor category formalism described in \cite{barkeshli_symmetry_2014}. 

In topological codes of dimension greater than 2 the braiding relations of defects and excitations are very poorly understood. Braiding in general is a more complicated concept in these higher dimensions and must be performed with non-pointlike objects to be non-trivial. Recent work shows that defects can be constructed in higher dimensions that reproduce the braiding relations of Ising anyons \cite{webster_fault-tolerant_2019} and it is also known that using domain walls and puncture encodings we can perform a braided version of any single-qubit transversal gate within a code. A rigorous examination of such braiding schemes should be possible using the full G-crossed category formalism since puncture encodings utilise the fact that some of the natural anyons of the code can condense at the puncture's boundary. Passing these punctures through a domain wall should therefore be equivalent to braiding these natural anyons with a twist and modifying the anyon's charge in the process. It remains to be seen whether or not it is possible to implement a braided non-Clifford gate that is not also a transversal gate of the code. 

The results of this work could also be extended to qudit codes, provided the anyon models of these codes possess symmetries. The twists that arise in such cases may again allow us to obtain non-Abelian anyons in codes where all the natural anyons are Abelian, but in cases where the natural anyons are non-Abelian but non-universal there is the possibility that additional non-Abelian charges may be obtained which make these models universal. These models may therefore be a fruitful topic for future research. 

\section*{Acknowledgements}
The authors would like to thank M. Kesselring, J. Pachos, A. Farjami, S. Burton, R. Craigen and J. Bridgeman for helpful discussions. In particular we would like to thank R. Craigen for his assistance regarding symmetric Hadamard matrices and J. Bridgeman for pointing out the connection to Tambara-Yamagami categories. TRS acknowledges support from University College London and the Engineering and Physical Sciences Research Council [grant number EP/L015242/1]. DEB acknowledges support from the Engineering and Physical Sciences Research Council QCDA project [grant number EP/R043647/1]. 

\section*{Appendices}
\appendix
\section{Derivation of F Matrices}
Since we will be using the pentagon equation (\ref{pentagoneq}) extensively in this section we quote it again here
\begin{equation}
    \label{pentagoneq2}
    (F_{12c}^5)_a^d(F_{a34}^5)_b^c = \sum_e(F_{234}^d)_e^c(F_{1e4}^5)_b^d(F_{123}^b)_a^e
\end{equation}

\noindent $\beta$ is non-Abelian so $(F_{\beta\beta\beta}^{\beta})$ is a matrix which we will henceforth refer to as $\bm{F}$. If we let $\bm{1} = \alpha_0$ then the elements of $\bm{F}$ are $F_{ij} = (F_{\beta\beta\beta}^{\beta})_{\alpha_i}^{\alpha_j}$.  All other $(F_{pqr}^s)_i^j$ are phases. We note some important properties of these phases and their inverses:

\begin{enumerate}
    \item Every phase $(F_{pqr}^s)_i^j$ has an inverse $(F_{rqp}^s)_j^i$ since changes of fusion order are always reversible. 
    \item $(F_{pqr}^s)_i^j$ where $p$,$q$ or $r = \bm{1}$ are trivial reorderings and correspond to a phase of $1$. 
    \item Phases of the type $(F_{pqp}^s)_i^i$ are self-inverse and so have value either $1$ or $-1$. 
\end{enumerate}

We begin with the case where $1,2,3,4=\beta$, $5=\bm{1}$, $b=d=\beta$, $a = \alpha_i$ and $c = \alpha_j$. We obtain constraints
\begin{equation}
    \label{mateq}
        (F_{\beta\beta\alpha_j}^{\bm{1}})_{\alpha_i}^{\beta}(F_{\alpha_i\beta\beta}^{\bm{1}})_{\beta}^{\alpha_j} = 
        \sum_{x=0}^{2^k-1}F_{xj}(F_{\beta\alpha_x\beta}^{\bm{1}})_{\beta}^{\beta}F_{ix}
\end{equation}

In the case that $i=j$ the LHS of this equation is equal to 1 by property 1 as listed above. If $i \neq j$ the LHS is equal to 0 since $\alpha_i \times \alpha_j = \bm{1}$ is only possible when $i=j$. If we write $(F_{\beta\alpha_x\beta}^{\bm{1}})_{\beta}^{\beta} = \theta_x $ and sum over repeated indices then we can rewrite (\ref{mateq}) as 
\begin{equation}
    \label{mateq2}
    F_{ix}F'_{xj} = \delta_{ij}
\end{equation}

\noindent where each element $F'_{ij} = \theta_i \cdot F_{ij}$ (we do not sum over the repeated index here). The LHS of (\ref{mateq2}) is the matrix $\bm{F}$ and its inverse. $\bm{F}$ is unitary so we must have that
\begin{equation}
    F_{ij} = (F'_{ij})^\dag.
\end{equation}

By property 3 $\theta_i = \pm 1$ and by property 2 $\theta_0 = 1$. Thus 
\begin{equation}
    F_{i0} = (F'_{0i})^* = (\theta_0 \cdot F_{0i})^* = (F_{0i})^*
\end{equation}

\noindent and 
\begin{equation}
    F_{0j} = (F'_{j0})^* = (\theta_j \cdot F_{j0})^* = (\pm F_{j0})^*
\end{equation}

\noindent and we can combine these to show
\begin{equation}
    F_{0x} = (\pm F_{x0})^* = (\pm(F_{0x})^*)^* = \pm F_{0x}.
\end{equation}

Thus if $\theta_x = -1$ we must have $F_{0x} = F_{x0} = 0$ for all $x\geq0$.

The next configurations of the pentagon equation that we consider are $1=\alpha_x$, $2,3,4=\beta$ and $2=\alpha_x$, $1,3,4=\beta$. The first of these yields constraints
\begin{equation}
    \label{ixx}
    F_{ij} = \frac{(F_{\alpha_x(\alpha_i \times \alpha_x)\beta}^{\beta})_{\alpha_i}^{\beta}
                         (F_{\alpha_x\beta\beta)}^{\alpha_i})_{\beta}^{\alpha_i \times \alpha_x}}
                        {(F_{\alpha_x\beta\alpha_j}^{\beta})_{\beta}^{\beta}}
                        F_{(i \times x)j}
\end{equation}

\noindent where $F_{(i \times x)j} = (F_{\beta\beta\beta}^{\beta})_{\alpha_i \times \alpha_x}^{\alpha_j}$. The second configuration yields
\begin{equation}
    \label{jxx}
    F_{ij} = \frac{(F_{\alpha_x\beta\beta}^{\alpha_j \times \alpha_x})_{\beta}^{\alpha_j}
                         (F_{\beta\alpha_x\beta}^{\alpha_i})_{\beta}^{\beta}}
                        {(F_{\beta\alpha_x\alpha_j}^{\beta})_{\beta}^{\alpha_j \times \alpha_x}}
                        F_{i(j \times x)}
\end{equation}

From (\ref{ixx}) we have that
\begin{equation}
    \label{ij0}
    F_{00} = (F_{\alpha_x\alpha_x\beta}^{\beta})_{\bm{1}}^{\beta}
                   (F_{\alpha_x\beta\beta}^{\bm{1}})_{\beta}^{\alpha_x}
                   F_{x0}
\end{equation}

\noindent and from (\ref{jxx})
\begin{equation}
    \label{j0}
    F_{i0} = (F_{\alpha_{x'}\beta\beta}^{\alpha_{x'}})_{\beta}^{\bm{1}}
                   (F_{\beta\alpha_{x'}\beta}^{\alpha_i})_{\beta}^{\beta}
                   F_{ix'}
\end{equation}

Setting $x=i$ and $x'=j$ we can combine these equations to obtain
\begin{equation}
    \label{oneij}
   F_{ij} = (F_{\alpha_j\alpha_j\beta}^{\beta})_{\bm{1}}^{\beta}(F_{\alpha_j\beta\beta}^{\bm{1}})_{\beta}^{\alpha_j}(F_{\alpha_i\beta\beta}^{\alpha_i})_{\beta}^{\bm{1}}(F_{\beta\alpha_i\beta}^{\alpha_j})_{\beta}^{\beta}F_{00}
\end{equation}

Thus all elements $F_{ij}$ have magnitude equivalent to that of $F_{00}$ and so the only way for $\theta_x = -1$ is to have $F_{00} = 0$ but this contradicts the constraint that $F_{ix}F'_{xi} = 1$ since all the terms in this sum would be 0. Thus all $\theta_x = 1$ and $\bm{F}$ must be Hermitian. Additionally, the magnitude of all elements in the matrix must be $1/\sqrt{2^k}$ and since the diagonal elements must be real $F_{00} = \pm 1/\sqrt{2^k}$.  

Setting
\begin{equation}
    \begin{split}
        &\phi_{ij} = (F_{\beta\alpha_i\beta}^{\alpha_j})_{\beta}^{\beta}, \\
        &f_{i0} = (F_{\alpha_i\beta\beta}^{\alpha_i})_{\beta}^{\bm{1}}, \\
        &f_{0j} = (F_{\alpha_j\alpha_j\beta}^{\beta})_{\bm{1}}^{\beta}
             (F_{\alpha_j\beta\beta}^{\bm{1}})_{\beta}^{\alpha_j}
    \end{split}
\end{equation}

\noindent we can rewrite (\ref{oneij}) as 
\begin{equation}
    \label{twoij}
    F_{ij} = \pm\frac{\phi_{ij}f_{i0}f_{0j}}{\sqrt{2^k}}.
\end{equation}

By property 3 all $\phi_{ij} = \pm 1$ (with $\phi_{i0} = \phi_{0j} = 1$ by property 2) while $f_{i0}$ can be any phase. We also have that $f_{0j} = (f_{j0})^*$ which can be verified by considering the five-anyon fusion tree

\begin{center}
    \begin{tikzpicture}

        \node [above] at (0,4) {$\alpha_i$};
        \node [above] at (1,4) {$\beta$};
        \node [above] at (2,4) {$\beta$};
        \node [above] at (3,4) {$\beta$};
        \node [above] at (4,4) {$\beta$};

        \draw (0,4) -- (2,0);
        \draw (1,4) -- (1.5,3);
        \draw (2,4) -- (1.5,3);
        \draw (3,4) -- (3.5,3);
        \draw (4,4) -- (2,0);

        \node [right] at (1.5,3) {$\alpha_i$};
        \node [right] at (3.5,3) {$\bm{1}$};

        \draw (1.5,3) -- (1,2);

        \node [left] at (1,2) {$\bm{1}$};

        \draw (2,0) -- (2,-1);

        \node [below] at (2,-1) {$\bm{1}$};

    \end{tikzpicture}
\end{center}

\noindent and observing that the F-move sequences $(F_{\alpha_i\alpha_i\beta}^{\beta})_{\bm{1}}^{\beta}$$(F_{\beta\beta\bm{1}}^{\bm{1}})_{\bm{1}}^{\beta}$ and $(F_{\beta\beta\alpha_i}^{\bm{1}})_{\alpha_i}^{\beta}$$(F_{\beta\beta\alpha_i}^{\alpha_i})_{\bm{1}}^{\beta}$$(F_{\alpha_i\alpha_i\bm{1}}^{\bm{1}})_{\bm{1}}^{\alpha_i}$ both map this tree to 

\begin{center}
    \begin{tikzpicture}

        \node [above] at (0,4) {$\alpha_i$};
        \node [above] at (1,4) {$\beta$};
        \node [above] at (2,4) {$\beta$};
        \node [above] at (3,4) {$\beta$};
        \node [above] at (4,4) {$\beta$};

        \draw (0,4) -- (2,0);
        \draw (1,4) -- (1.5,3);
        \draw (2,4) -- (1.5,3);
        \draw (3,4) -- (1.5,1);
        \draw (4,4) -- (2,0);

        \node [right] at (1.5,3) {$\alpha_i$};

        \draw (1.5,3) -- (2,2);

        \node [right] at (2,2) {$\beta$};

        \node [left] at (1.5,1) {$\beta$};
        
        \draw (2,0) -- (2,-1);

        \node [below] at (2,-1) {$\bm{1}$};

    \end{tikzpicture}.
\end{center}

Equating these and eliminating trivial terms using property 2 we get
\begin{equation}
    (F_{\alpha_i\alpha_i\beta}^{\beta})_{\bm{1}}^{\beta} = (F_{\beta\beta\alpha_i}^{\bm{1}})_{\alpha_i}^{\beta}(F_{\beta\beta\alpha_i}^{\alpha_i})_{\bm{1}}^{\beta}
\end{equation}

\noindent which can be rewritten as 
\begin{equation}
    (F_{\alpha_i\alpha_i\beta}^{\beta})_{\bm{1}}^{\beta}(F_{\alpha_i\beta\beta}^{\bm{1}})_{\beta}^{\alpha_i} = ((F_{\alpha_i\beta\beta}^{\alpha_i})_{\beta}^{\bm{1}})^{-1}
\end{equation}

\noindent using property 1.

We can then write $\bm{F}$ as the Hadamard product of a matrix of $f$s ($\bm{f}$) and a matrix of $\phi$s ($\bm{\phi}$) multiplied by $1/\sqrt{2^k}$ 
\begin{equation}
    \bm{F} = \pm\frac{1}{\sqrt{2^k}} (\bm{\phi} \circ \bm{f}).
\end{equation}

\noindent $\bm{f}$ is Hermitian and so $\bm{\phi}$ must also be Hermitian. The multiplication of $\bm{F}$ by itself gives
\begin{equation}
    \begin{split}
        F_{ix}F_{xj} &= \frac{\phi_{ix}f_{i0}f_{0x}}{\sqrt{2^k}}\frac{\phi_{xj}f_{x0}f_{0j}}{\sqrt{2^k}}  \\
                     &= \frac{1}{2^k} \cdot \phi_{ix}\phi_{xj} \cdot f_{i0}f_{0j} \\
                     &= \delta_{ij}
    \end{split}
\end{equation}

\noindent where we have used the fact that $f_{0x}f_{x0}=1$. The final equivalence implies 
\begin{equation}
    \frac{1}{2^k}\bm{\phi}^2 \circ \bm{f} = \bm{I}
\end{equation}

\noindent where $\bm{I}$ is the $2^k \times 2^k$ identity matrix. The diagonal elements of $\bm{f}$ are all 1 (since $f_{0i} = (f_{i0})^*$) so we must have that $\bm{\phi}^2 = 2^k\bm{I}$. Thus $\bm{\phi}$ is a symmetric Hadamard matrix \cite{hedayat_hadamard_1978}. 

$\bm{\phi}$ has only a finite number of solutions while $\bm{f}$ has an infinite number. The obvious interpretation of these two matrices is that different $\bm{\phi}$ correspond to different anyon models (with the discreteness of these solutions consistent with Ocneanu rigidity \cite{kitaev_anyons_2006}), while the different $\bm{f}$ correspond to a choice of gauge. We can see that we cannot transform between solutions of $\bm{\phi}$ by changes to $\bm{f}$ by observing that $\bm{f}$ is completely characterised by the values of $f_{i0}$, whereas $\phi_{i0}$ are always $1$, so changes to $\bm{f}$ are always reflected in the first row of $F_{\beta\beta\beta}^{\beta}$ while changes to $\bm{\phi}$ are not. We also show in appendix \ref{Rapp} that the braiding matrices of the models depend only on $\bm{\phi}$ and not on $\bm{f}$. Thus we can make the gauge choice that all $\bm{f}=1$ so
\begin{equation}
    \bm{F} = \pm\frac{1}{\sqrt{2^k}}\bm{\phi}.
\end{equation}

Note that instead of (\ref{ij0}) and (\ref{j0}) we could have obtained 
\begin{equation}
    F_{0j} = (F_{\alpha_x\alpha_x\beta}^{\beta})_{\bm{1}}^{\beta}(F_{\alpha_x\beta\beta}^{\bm{1}})_{\beta}^{\alpha_x}(F_{\alpha_j\beta\alpha_x}^{\beta})_{\beta}^{\beta}F_{xj}
\end{equation}

\noindent from (\ref{ixx}) and
\begin{equation}
    F_{00} = (F_{\alpha_x'\beta\beta}^{\alpha_x'})_{\beta}^{\bm{1}}F_{0x'}
\end{equation}

\noindent from (\ref{jxx}). Setting $x=i$ and $x'=j$ and combining these as before gives
\begin{equation}
    F_{ij} = (F_{\beta\beta\alpha_i}^{\bm{1}})_{\alpha_i}^{\beta}(F_{\beta\alpha_i\alpha_i}^{\beta})_{\beta}^{\bm{1}}(F_{\beta\beta\alpha_j}^{\alpha_j})_{\bm{1}}^{\beta}(F_{\alpha_i\beta\alpha_j}^{\beta})_{\beta}^{\beta}F_{00}.
\end{equation}

Using property 1 we can see that the first three terms are equal to $(f_{j0}f_{0i})^{-1}$ which is equal to 1 due to our choice of gauge. Thus we have that 
\begin{equation}
    \frac{1}{\sqrt{2^k}}\phi_{ij} = \frac{1}{\sqrt{2^k}}(F_{\alpha_i\beta\alpha_j}^{\beta})_{\beta}^{\beta}
\end{equation}

\noindent so $\phi_{ij} = (F_{\beta\alpha_i\beta}^{\alpha_j})_{\beta}^{\beta} = (F_{\alpha_i\beta\alpha_j}^{\beta})_{\beta}^{\beta}$ with this gauge choice. 

\label{Fapp}
\section{Column Permutations of Symmetric Hadamard Matrices}
In this appendix we show that column permutations that preserve the symmetry of a symmetric $2^k \times 2^k$ Hadamard matrix must alter the trace by either $0$ or $\pm2^k$. Consider swapping the second and third columns of the following matrix from (\ref{k2fmat}). This matrix corresponds to $H_1 \otimes H_1$ and this column swap will result in the matrix from (\ref{k2fmat}) with trace 4.

\begin{center}
\begin{tikzpicture}
    \matrix (m) [
        matrix of math nodes, 
        nodes in empty cells, 
        left delimiter={(},
        right delimiter={)},
        nodes={text width={width(-1)},align=right}]
    {
        1 & 1 & 1 & 1 \\
        1 & -1 & 1 & -1 \\
        1 & 1 & -1 & -1 \\
        1 & -1 & -1 & 1 \\
    };
    \draw (m-1-2.south west) rectangle (m-1-3.north east);
    \draw (m-3-2.south west) rectangle (m-2-3.north east);
    \draw (m-4-2.south west) rectangle (m-4-3.north east);
\end{tikzpicture}
\end{center}

We have divided this pair of columns into three sections. Any changes to the top or bottom (off-diagonal) sections will result in a non-symmetric matrix but the central (diagonal) section can be modified while preserving symmetry. Because these two columns are identical in the off-diagonal sections and the diagonal section is symmetric both before and after the exchange the symmetry of the overall matrix is preserved. However, for $k>2$ such exchanges cannot be symmetry preserving because the diagonal section will always contain 2 elements from each column while the off-diagonal sections will contain the other $2^k-4$. In order for the columns of the matrix to be orthogonal each pair of columns must match in exactly half their entries so for matrices larger than $4\times4$ it is impossible to exchange two columns in such a way that the off-diagonal sections are unchanged. Instead, we must exchange sets of columns. Consider a general $2^k \times 2^k$ matrix broken into $2^{k-2} \times 2^{k-2}$ blocks as follows

\begin{center}
\begin{tikzpicture}
    \matrix (m) [
        matrix of math nodes, 
        nodes in empty cells, 
        left delimiter={(},
        right delimiter={)}]
    {
        & E & F & \\
        & A & B & \\
        & C & D & \\
        & G & H & \\
    };
    \draw (m-1-2.south west) rectangle (m-1-3.north east);
    \draw (m-3-2.south west) rectangle (m-2-3.north east);
    \draw (m-4-2.south west) rectangle (m-4-3.north east);
\end{tikzpicture}
\end{center}

\noindent where we have once again marked diagonal and off-diagonal sections. 

\textbf{Lemma 1:} \textit{The exchange of $EACG$ and $FBDH$ can only preserve the symmetry of the matrix if the trace of the diagonal section is negated by the exchange}

We can only exchange $EACG$ and $FBDH$ if $E=F$ and $G=H$. Additionally we must have that $A=-B$ and $C=-D$ or the columns of the matrix would not be orthogonal. Thus the trace of the diagonal section is negated by this exchange. $\square$

\textbf{Lemma 2:} \textit{The exchange of $EACG$ and $FBDH$ can only preserve the symmetry of the matrix if $A$ is a symmetric Hadamard matrix}

For the overall matrix to be symmetric both before and after the exchange we require that $B^T=C$ and $A^T=D$ and therefore $A=-B=-C^T=D^T$ and the entire diagonal section is determined by $A$. 

$A$ is must be symmetric for the overall matrix to be symmetric.

To show that the columns of $A$ are orthogonal we consider two columns, $c_1$ and $c_2$, of our full $2^k \times 2^k$ matrix such that both columns pass through $A$. We note that if this pair of columns have $m$ matched elements within $A$ then they must have $2m$ matched elements within the diagonal section and $2^{k-1}-2m$ matched elements in the off-diagonal sections (since exactly half the elements of each column must match). We now consider the pair $c_1$ and $-c_2$ where $-c_2$ is the column passing through $B$ that is equal to $-1\times c_2$ within the diagonal section and identical to $c_2$ outside of this section (i.e. the column we wish to exchange $c_2$ with). $c_1$ and $-c_2$ have $2^{k-1}-2m$ matched elements within the diagonal section and so must have $2m$ matched elements in the off-diagonal sections. Since $c_2$ and $-c_2$ are identical in the off-diagonal sections we have that $2m = 2^{k-1}-2m$ and $m = 2^{k-3}$. Thus for each pair of columns in $A$ half of the elements are matched and the other half are unmatched and the columns of $A$ are orthogonal.   $\square$

Since $A$ is a symmetric Hadamard matrix of order $2^{k-2}$ it can be partitioned into blocks such that $A'$ is a symmetric Hadamard matrix of order $2^{k-4}$. This process can be repeated recursively, eventually terminating when we arrive at a matrix of order 1 (for odd $k$) or order 2 (for even $k$). The possible symmetric Hadamard matrices with these orders are (up to a possible global phase of -1) (\ref{H1}) and (\ref{k2fmat}). These matrices are restricted to have trace=0 and trace=0 or 4 respectively, and going back up the chain of recursion we see that Tr$(A)=0$ for odd $k$ and Tr$(A)=0,\pm 2^{k-1}$ for even $k$. The trace of the central section is then $2\textrm{Tr}(A)$ and by Lemma 1 a symmetry preserving exchange of columns must negate this trace, changing the trace of the overall matrix by either $0$ for odd $k$ or $0,2^k$ for even $k$. 

So far we have only considered swapping columns in a very specific arrangement, but any desired set of column swaps can be rewritten in this form such that it is apparent that the same constraints apply. This is achieved by applying column permutations such that the columns we wish to swap are correctly arranged at the centre of the matrix and then applying a matching set of row permutations (since matching column and row permutations preserve the symmetry of the matrix). Following the exchange of column blocks as described above we apply the same set of row and column permutations again. By considering how permutations transform under conjugation and the fact that row and column permutations commute we can see that this operation is equivalent to exchanging the columns without first moving them to the center. 

\label{permuteapp}
\section{Derivation of R Matrices}
The R-matrix is defined from the hexagon equation
\begin{equation}
    R_{13}^c(F_{213}^4)_a^cR_{12}^a = \sum_b(F_{231}^4)_b^cR_{1b}^4(F_{123}^4)_a^b
\end{equation}

When we set $1,2,3,4 = \beta$, $a,c = \alpha_i$ this gives:
\begin{equation}
    R_{\beta\beta}^{\alpha_j}F_{ij}R_{\beta\beta}^{\alpha_i} = \sum_{b = 0}^{2^k}F_{bj}R_{\beta\alpha_b}^{\beta}F_{ib}
\end{equation}

\noindent and from (\ref{twoij})
\begin{equation}
    \begin{split}
        \pm\frac{\phi_{ij}f_{i0}f_{0j}}{\sqrt{2^k}}&R_{\beta\beta}^{\alpha_i}R_{\beta\beta}^{\alpha_j}
        =  \\
        &\sum_{b = 0}^{2^k}\frac{\phi_{bj}f_{b0}f_{0j}\phi_{ib}f_{i0}f_{0b}R_{\beta \alpha_b}^{\beta}}{2^k}
    \end{split}
\end{equation}
\begin{equation}
    \label{R1}
    \pm\frac{\phi_{ij}R_{\beta\beta}^{\alpha_i}R_{\beta\beta}^{\alpha_j}}{\sqrt{2^k}}
    =
    \sum_{b = 0}^{2^k}\frac{\phi_{ib}\phi_{bj}R_{\beta \alpha_b}^{\beta}}{2^k}
\end{equation}

Setting $1=\alpha_i$ and $4=\alpha_j$ we instead obtain
\begin{equation}
    \label{R5}
    \phi_{ij}(R_{\alpha_i\beta}^{\beta})^2 = R_{\alpha_i(\alpha_i \times \alpha_j)}^{\alpha_j}
\end{equation}

So we see that the braiding relations are dependent on $\bm{\phi}$ but not on ${\bf f}$ as we would expect.

Consider the case in (\ref{R1}) where $i=j$. We have that
\begin{equation}
    \frac{\phi_{ii}(R_{\beta\beta}^{\alpha_i})^2}{\sqrt{2^k}} = 
    \pm\sum_{b}\frac{R_{\beta\alpha_b}^{\beta}}{2^k}.
\end{equation}

The RHS is independent of $i$ so 
\begin{equation}
    \label{R2}
    (R_{\beta\beta}^{\alpha_0})^2 = \phi_{ii}(R_{\beta\beta}^{\alpha_i})^2.
\end{equation}

Additionally, if we set $i=0$ and sum over $j$
\begin{equation}
    \pm\sum_j\frac{R_{\beta\beta}^{\alpha_0}R_{\beta\beta}^{\alpha_j}}{\sqrt{2^k}} =
    \sum_j\sum_b\frac{\phi_{bj}R_{\beta\alpha_b}^{\beta}}{2^k}
\end{equation}

\noindent all terms on the RHS cancel except for those where $b=0$ which sum to $2^kR_{\beta\alpha_0}^{\beta}$ (since all rows/columns of $\bm{\phi}$ except for the first contain an equal number of $1$s and $-1$s), so 
\begin{equation}
    \pm\sum_j\frac{R_{\beta\beta}^{\alpha_0}R_{\beta\beta}^{\alpha_j}}{\sqrt{2^k}} =
    R_{\beta\alpha_0}^{\beta} = 1
\end{equation}

\noindent since $R_{\beta\alpha_0}^{\beta}$ just describes braiding with the vacuum and is therefore trivial. Using (\ref{R2}) we can rewrite this as
\begin{equation}
    \label{R3}
    \pm\frac{(R_{\beta\beta}^{\alpha_0})^2(1 \pm \sqrt{\phi_{11}} \pm ... \pm \sqrt{\phi_{n_kn_k}})}{\sqrt{2^k}} = 1
\end{equation}

The $\pm$ in the sum do not all need to be the same, but they must be chosen such that $|R_{\beta\beta}^{\alpha_0}|^2=1$ or the matrix $R_{\beta\beta}$ would not be unitary. 

The number of $+1$ ($-1$) terms in the diagonal of $\bm{\phi}$ tells us the number of $\pm 1$ ($\pm i$) terms in the sum of (\ref{R3}) which we can rewrite as 
\begin{equation}
    \label{R4}
    \pm\frac{(R_{\beta\beta}^{\alpha_0})^2(a + ib)}{\sqrt{2^k}} = 1.
\end{equation}

In order for $|R_{\beta\beta}^{\alpha_0}|^2=1$ we require that $a^2 + b^2 = 2^k$. For even $k$ this means that either $a=\pm2^{k/2},b=0$ or $a=0,b=\pm2^{k/2}$. For odd $k$ we have that $a = \pm b = \pm2^{(k-1)/2}$. The proofs are as follows:

Consider a right-angled triangle with sides $a \leq b < c$ opposed by angles $A,B,C$.

\begin{itemize}
    \item \textbf{Even:} $\sqrt{2^k} = 2^{k/2}$ is an integer so from $a^2 +b^2 = 2^k$ we assume the existence of a Pythagorean triple $(|a|,|b|,2^{k/2})$. $a$ and $b$ must have the same parity for the sum of their squares to be even. If they are both odd then this triple is primitive since $|a|$ and $|b|$ have no even factors and $2^{k/2}$ has no odd factors. If they are both even then there must be some associated primitive triple $(|a|/2^x,|b|/2^x,2^{k/2-x})$ where the first two elements are both odd. All primitive triples can be constructed using Euclid's formula
        \begin{equation}
            a = m^2 - n^2, ~~ b = 2mn, ~~ c = m^2 + n^2
        \end{equation}

        where $m$ and $n$ are a pair of coprime integers, one of which is even. However, this means that $c$ is odd, giving a contradiction. Thus this primitive triple does not exist and neither does the triple $(|a|,|b|,2^{k/2})$. The only remaining solutions to the equation $a^2 + b^2 = 2^k$ are $a=\pm2^{k/2},b=0$ and $a=0,b=\pm2^{k/2}$
    \item \textbf{Odd:} $\sqrt{2^k} = 2^{k/2} = 2^{(k-1)/2}\sqrt{2}$ where $2^{(k-1)/2}$ is an integer power of two. Given $c = 2^{(k-1)/2}\sqrt{2}$ we have that $a = 2^{(k-1)/2}\sqrt{2}\textrm{sin}(A)$ and $b = 2^{(k-1)/2}\sqrt{2}\textrm{sin}(B)$. We require that both $a$ and $b$ are integers and thus $\textrm{sin}(A)$ and $\textrm{sin}(B)$ must both be integer multiples of $1/\sqrt{2}$. Thus $\textrm{sin}(A) = \pm \textrm{sin}(B) = \pm 1/\sqrt{2}$ and $a = \pm b = \pm2^{(k-1)/2}$. 
\end{itemize}

Substituting these solutions into (\ref{R4}) we have 
\begin{equation}
    \label{Req1}
    \pm(R_{\beta\beta}^{\alpha_0})^2 = 1 ~~ \textrm{and} ~~ \pm i(R_{\beta\beta}^{\alpha_0})^2 = 1
\end{equation}

\noindent for even $k$ and 
\begin{equation}
    \label{Req2}
    \pm\frac{(R_{\beta\beta}^{\alpha_0})^2(1 \pm i)}{\sqrt{2}} = 1
\end{equation}

\noindent for odd $k$. 

Finally, we note that by setting $i=j$ in (\ref{R5}) we can show that 
\begin{equation}
    R_{\alpha_i\beta_k}^{\beta_k} = \pm\sqrt{\phi_{ii}}\sqrt{R_{\alpha_i{\bf 1}}^{\alpha_i}} = \pm\sqrt{\phi_{ii}}
\end{equation}

\noindent where $R_{\alpha_i{\bf 1}}^{\alpha_i} = 1$ since braiding with the vacuum is trivial. Using this and instead setting $j=0$ we find 
\begin{equation}
    (R_{\alpha_i\beta_k}^{\beta_k})^2 = \phi_{ii} = R_{\alpha_i\alpha_i}^{\bf 1}.
\end{equation}

In other words the elements $\phi_{ii}$ tell us the self-exchange statistics of the charges $\alpha_i$.

\label{Rapp}

\bibliographystyle{unsrtnat}
\bibliography{references}

\end{document}